\documentclass[sigconf]{acmart}
\AtBeginDocument{%
  }

\usepackage{enumitem}
\usepackage{booktabs} % For professional looking tables
\usepackage{multirow} % To span rows
\usepackage{graphicx} % For resizebox
\usepackage{makecell} % For multi-line cells
\usepackage{caption}
\usepackage{quoting}
\usepackage{cleveref}
\usepackage{xspace}
\usepackage[linesnumbered,ruled,vlined]{algorithm2e}
\usepackage{array}        % for column formatting like 'C' in tabular
\usepackage{booktabs}     % optional, for better horizontal lines (not strictly needed here)
\usepackage{caption}      % for caption customization (if you use \caption)
\newcolumntype{P}{>{\raggedright\arraybackslash}m{0.98\linewidth}}
\newcolumntype{C}{>{\arraybackslash}m{1.\linewidth}}

\crefformat{section}{\S#2#1#3}
\crefformat{subsection}{\S#2#1#3}
\crefformat{subsubsection}{\S#2#1#3}

\setcopyright{cc}
\setcctype{by}
\copyrightyear{2026}
\acmYear{2026}
\acmConference[SIGIR '26]{Proceedings of the 49th International ACM SIGIR Conference on Research and Development in Information Retrieval}{July 20--24, 2026}{Melbourne, VIC, Australia.}
\acmBooktitle{Proceedings of the 49th International ACM SIGIR Conference on Research and Development in Information Retrieval (SIGIR '26), July 20--24, 2026, Melbourne, VIC, Australia}
\acmISBN{979-8-4007-2599-9/2026/07}
\acmDOI{10.1145/3805712.3809656}

\begin{document}

\title{Filling the Gaps: Selective Knowledge Augmentation for LLM Recommenders}

\author{Jaehyun Lee}
\orcid{0009-0002-1232-542X}
\affiliation{%
  \institution{Pohang University of Science and Technology}
  % \institution{POSTECH}
  \city{Pohang}
  \country{Republic of Korea}
}
\email{jminy8@postech.ac.kr}

\author{Sanghwan Jang}
\orcid{0009-0000-9856-491X}
\affiliation{%
  \institution{Pohang University of Science and Technology}
  % \institution{POSTECH}
  \city{Pohang}
  \country{Republic of Korea}
}
\email{s.jang@postech.ac.kr}

\author{SeongKu Kang}
\authornote{Corresponding authors.}
\orcid{0000-0001-5528-1426}
\affiliation{%
  \institution{Korea University}
  \city{Seoul}
  \country{Republic of Korea}
}
\email{seongkukang@korea.ac.kr}

\author{Hwanjo Yu}
\authornotemark[1]
\orcid{0000-0002-7510-0255}
\affiliation{%
  \institution{Pohang University of Science and Technology}
  % \institution{POSTECH}
  \city{Pohang}
  \country{Republic of Korea}
}
\email{hwanjoyu@postech.ac.kr}

\renewcommand{\shortauthors}{Lee et al.}

%%
%% The abstract is a short summary of the work to be presented in the
%% article.
\begin{abstract}

% Owing to their broad world knowledge, 
Large language models (LLMs) have recently emerged as powerful training-free recommenders.
However, their knowledge of individual items is inevitably uneven due to imbalanced information exposure during pretraining, a phenomenon we refer to as \textit{knowledge gap} problem.
To address this, most prior methods have employed a naive uniform augmentation that appends external information for every item in the input prompt.
However, this approach not only wastes limited context budget on redundant augmentation for well-known items but can also hinder the model's effective reasoning.
To this end, we propose \proposed (\textbf{K}nowledge-aware \textbf{S}elective \textbf{A}ugmentation with \textbf{C}omparative \textbf{K}nowledge \textbf{P}robing) to mitigate the knowledge gap problem. 
\proposed \textit{estimates} the LLM’s internal knowledge by evaluating its capability to capture collaborative relationships and \textit{selectively injects} additional information only where it is most needed.
By avoiding unnecessary augmentation for well-known items, \proposed focuses on items that benefit most from knowledge supplementation, thereby making more effective use of the context budget.
\proposed requires no fine-tuning step, and consistently improves both recommendation accuracy and context efficiency across four real-world datasets.
Our code is available at https://github.com/nowhyun/KnowSA\_CKP.

\end{abstract}

%%
%% The code below is generated by the tool at http://dl.acm.org/ccs.cfm.
%% Please copy and paste the code instead of the example below.
%%


%%
%% Keywords. The author(s) should pick words that accurately describe
%% the work being presented. Separate the keywords with commas.
% methods
\newcommand{\proposedtwo}{KnowSA\textsubscript{DKP}\xspace}
\newcommand{\proposed}{KnowSA\texorpdfstring{\textsubscript{CKP}}{\_CKP}\xspace}
\newcommand{\smallsection}[1]{{\vspace{0.05in} \noindent \bf {#1}}}


%% A "teaser" image appears between the author and affiliation
%% information and the body of the document, and typically spans the
%% page.

%%
%% This command processes the author and affiliation and title
%% information and builds the first part of the formatted document.
\maketitle

\section{Introduction}

% 1st paragraph: traditional recommender system - cold start/sparsity issue - generalability of LLM - traing free recommender
Recommender systems play a crucial role in helping users navigate the ever-growing landscape of digital content and products. 
Classical approaches such as matrix factorization \cite{MF-BPR} rely heavily on historical interaction data to infer user preferences.
However, these methods struggle in cold-start scenarios involving users or items with limited or no historical interactions, which severely restricts the system’s ability to generalize.
To overcome these challenges, recent studies have explored the use of large language models (LLMs) as knowledge-rich recommenders \cite{LLMRank-ECIR24, ALLMRec-KDD24, luo2024recranker, LLARA23, RLMRec-WWW24}. 
Pretrained on vast corpora, LLMs possess broad semantic understanding and factual knowledge about a wide range of entities and domains.
This allows them to serve as powerful \textit{training-free} recommenders without additional costly fine-tuning steps \cite{zero-shot-rec-COLING'25, LLMRank-ECIR24, zero-shot-Llamarec}.

% 2nd paragraph: knowledge gap problem of RecLLM - definition of gap - reason of gap - result of gap
Despite their potential, one critical challenge in using LLMs for recommendation is the imbalance in their parametric knowledge, a phenomenon we refer to as the \textit{knowledge gap}. 
As LLMs are pretrained using texts from the web, the amount of information they acquire about each item is inherently biased toward popular items with higher visibility online \cite{gap-ACL23, exposure-EMNLP'23, exposure2-EMNLP'23}, leaving the model with insufficient knowledge of long-tail items.
However, in the recommendation context, popularity alone serves as an imperfect indicator for this knowledge gap. 
This is because an effective recommendation requires more than merely knowing details about individual items; 
it demands capturing their \textit{collaborative patterns}, such as co-consumption behaviors in user histories and item-item relationships. 
Even among items with similar popularity, the extent to which an LLM acquire their relational context can differ substantially—one item might be associated with rich interaction patterns, while another exists in isolation.
Due to this uneven knowledge, LLMs often over-rely on knowledge-rich items and fail to provide personalized recommendations.
Despite its critical impact, little effort has been made to directly quantify and resolve this~knowledge~gap.

% 3rd paragraph: related works (model-level adaptation - cons -> prompt-level augmentation - cons)
While existing studies have not explicitly addressed this knowledge gap, many efforts can be seen as indirect attempts to alleviate it.
One straightforward approach is \textbf{model-level adaptation}, where the LLM is fine-tuned using user-item interaction data to directly learn collaborative patterns \cite{TallRec, CoLLM23, BinLLM-ACL24, LLARA23, LLM4SeqRec-RecSys23}. 
However, this approach incurs substantial computational costs and risks degrading the model's general capabilities, such as instruction following and explainability \cite{LoRA, yang2024unveiling}.
Consequently, recent research has increasingly focused on \textbf{prompt-level augmentation}, which enriches input prompts with item metadata or external knowledge in a training-free manner \cite{TransRec-KDD24, EXP3RT24, ItemRAG_arXiv25, LLM-TRSR-WWW24}.
However, there remains substantial room for improvement in this direction.
Most existing methods adopt \textit{uniform augmentation}, indiscriminately adding information for all items in the input prompt, regardless of how much the model already knows about each item.
This approach is suboptimal; it not only wastes the limited context budget by adding redundant information for known items, while simultaneously increasing the risk of performance degradation as LLMs struggle to interpret essential signals within excessively long contexts \cite{lostinthemiddle}.

% 4th paragraph: Selective의 필요성 제기 -> 기존 지식 측정 방식(Heuristics)의 실패 분석 -> 새로운 측정 방식의 필요성 도출
To resolve this inefficiency, one might consider adopting \textit{adaptive retrieval} strategies from the general NLP domain \cite{FLARE_EMNLP23, DRAGIN_ACL24}, which dynamically determine when to retrieve external information. 
However, directly applying these techniques to recommendation presents critical challenges.
Existing adaptive methods monitor queries sequentially during generation to detect knowledge deficiency.
Such an inference-time approach is ill-suited for recommendation scenarios, where minimizing inference latency.
Moreover, LLMs struggle to accurately discriminate knowledge gaps for multiple items simultaneously within a complex prompt.
To overcome these limitations, we propose a selective augmentation framework grounded in \textbf{offline knowledge estimation}. 
By pre-computing the knowledge necessity for each item, we can efficiently inject external information only where it is needed without incurring inference overhead.
This naturally raises the question of \textbf{how to quantify the degree of knowledge} that an LLM possesses for each item.
We begin by analyzing various knowledge proxies used to approximate the model's knowledge, ranging from heuristics like item popularity \cite{KRagRec25} to advanced signals such as generation likelihood \cite{Mink-ICLR24} and consistency metric \cite{EigValLaplacian, SeaKR}.
However, directly applying these general-purpose methods to recommendations yields suboptimal results.
As detailed in our analysis \cref{prelim_analysis}, these proxies correlate poorly with recommendation accuracy because they primarily assess only semantic familiarity (e.g., knowing what an item is).
They do not effectively capture the collaborative relationship between the user's history and the target item (e.g., co-consumption pattern), which is essential for recommendation.
These observations highlight the need for a recommendation-tailored knowledge scoring strategy that accounts for both semantic and collaborative aspects, enabling more targeted and effective prompt augmentation.

% 5th paragraph: 
In this work, we propose \proposed (\textbf{Know}ledge-aware \textbf{S}elective \textbf{A}ugmentation with \textbf{C}omparative \textbf{K}nowledge \textbf{P}robing), a training-free framework designed to mitigate item-level knowledge gaps in LLM-based recommendation. 
\proposed is designed to (1) estimate the LLM’s knowledge for each item, and (2) selectively inject additional information only where it is most needed.
To this end, we introduce a new knowledge scoring strategy, called CKP, which quantifies the model's capability to comparatively rank items based on collaborative patterns.
Guided by this score, we employ a personalized augmentation strategy that selectively enriches knowledge-poor items with relevant reference anchors to bridge the semantic gap, effectively activating latent knowledge already encoded in the model.
Unlike prior methods that uniformly augment all items, \proposed makes more efficient use of the context budget by focusing on items that benefit most from supplementation.

The paper makes the following key contributions:
\begin{itemize}[leftmargin=*] \vspace{-\topsep}
    \item \textbf{Problem.} We formally present the necessity and difficulty of resolving the knowledge gap problem in LLM-based recommendation, which remains less explored in the previous literature.

    \item \textbf{Analysis.} We provide a comprehensive analysis of various knowledge proxies for estimating an LLM's knowledge in recommendation tasks, shedding light on the design of an effective solution.

    \item \textbf{Algorithm.} We propose \proposed, which is equipped with new knowledge scoring and augmentation strategies tailored to recommendation tasks. 
    As a plug-and-play framework, it can be flexibly applied to various LLMs.
    
    \item \textbf{Experiments.} Extensive experiments show that \proposed consistently improves recommendation quality in both accuracy and diversity, with negligible additional latency.

\end{itemize}\vspace{-\topsep}

\section{Preliminaries}
\label{subsec:definition}
\noindent
\textbf{Notations.}
Let the dataset be represented as $\mathcal{D} = (\mathcal{U}, \mathcal{I}, \mathcal{T}, \mathcal{H})$, where $\mathcal{U}$, $\mathcal{I}$, $\mathcal{T}$, and $\mathcal{H}$ represent the sets of users, items, item attribute texts, and user interaction histories, respectively. 
For each user $u \in \mathcal{U}$, the interaction sequence is denoted as $H^u = (i^u_1, i^u_2, \dots, i^u_{|H^u|}) \in \mathcal{H}$, where $|H^u|$ is the length of user $u$'s interaction sequence, and $i^u_k \in \mathcal{I}$ is the $k$-th item user $u$ interacted with chronologically. 
Each item $i \in \mathcal{I}$ is associated with textual features, represented as $(t^i, {a^i}) \in \mathcal{T}$, where $t^i$ is the item's title and ${a^i}$ are additional textual attributes (e.g., genre, developer, and description).

\smallsection{Recommendation with an LLM ranker.}
We adopt the standard two-stage recommendation \cite{covington2016deep, kang2019candidate}, consisting of \textit{candidate generation} followed by \textit{ranking}.  
In the first stage, a small set of candidate items $C^u = \{i_1^u, i_2^u, \dots, i_m^u\}$ is retrieved from the full item set $\mathcal{I}$ using lightweight models, where $m \ll |\mathcal{I}|$.
In the second stage, we employ LLMs as \textit{training-free rankers}~\cite{LLMRank-ECIR24, zero-shot-setting-ICLR'22}, which orders the candidate items based on its parametric knowledge and the input prompt. 
The prompt is constructed using the user's history $H^u$, the candidate set $C^u$, and their associated features from~$\mathcal{T}$.

\smallsection{Problem Definition.}
We follow the two-stage recommendation, where the LLM is leveraged as a training-free ranker.
Each item is represented by its title as the simplest form of input, and can be further augmented using additional attributes \cite{ReLLa-WWW'24, TransRec-KDD24}.
Our goal is to develop a plug-and-play framework that mitigates the knowledge gap problem by selectively augmenting input prompts. % without fine-tuning.

We seek to \textit{estimate} the LLM’s internal knowledge for each item and \textit{selectively inject} additional information only where it is most needed.
By avoiding redundant augmentation for well-known items, the framework can utilize the context budget more effectively by allocating it to items that benefit most from knowledge supplement.

\begin{figure}[b]
\centering
\includegraphics[width=0.9\linewidth]{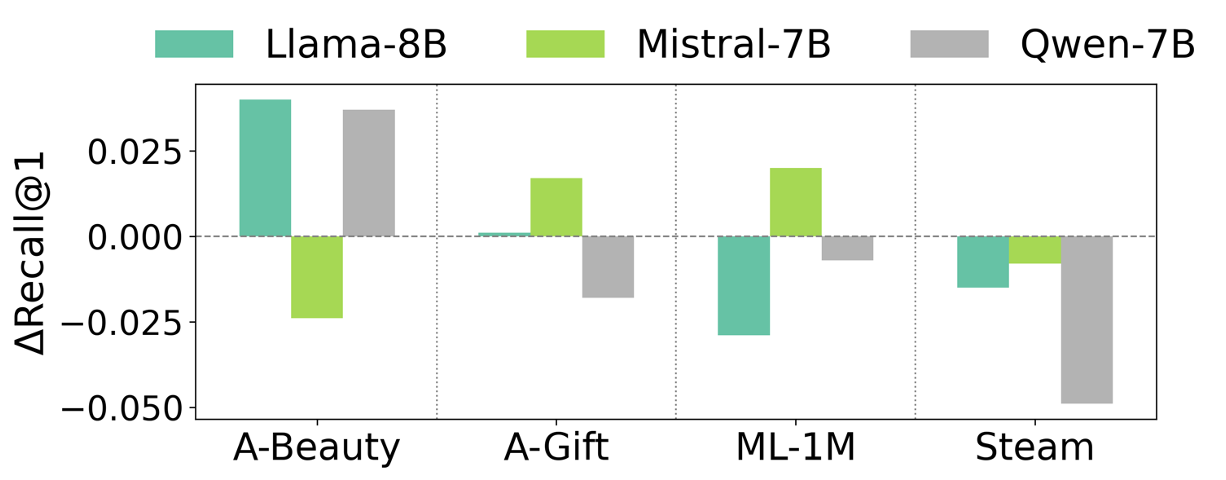}
\Description{}
% \vspace{-4mm}
\caption{
Recommendation performance change with uniform augmentation for all items. 
Indiscriminate augmentation for all items can rather hurt performance.
}
\label{fig:ff-gain}
% \vspace{-1mm}
\end{figure}

\begin{figure*}[t]
\centering
\begin{minipage}{0.32\linewidth}
    \centering
    \includegraphics[width=\linewidth]{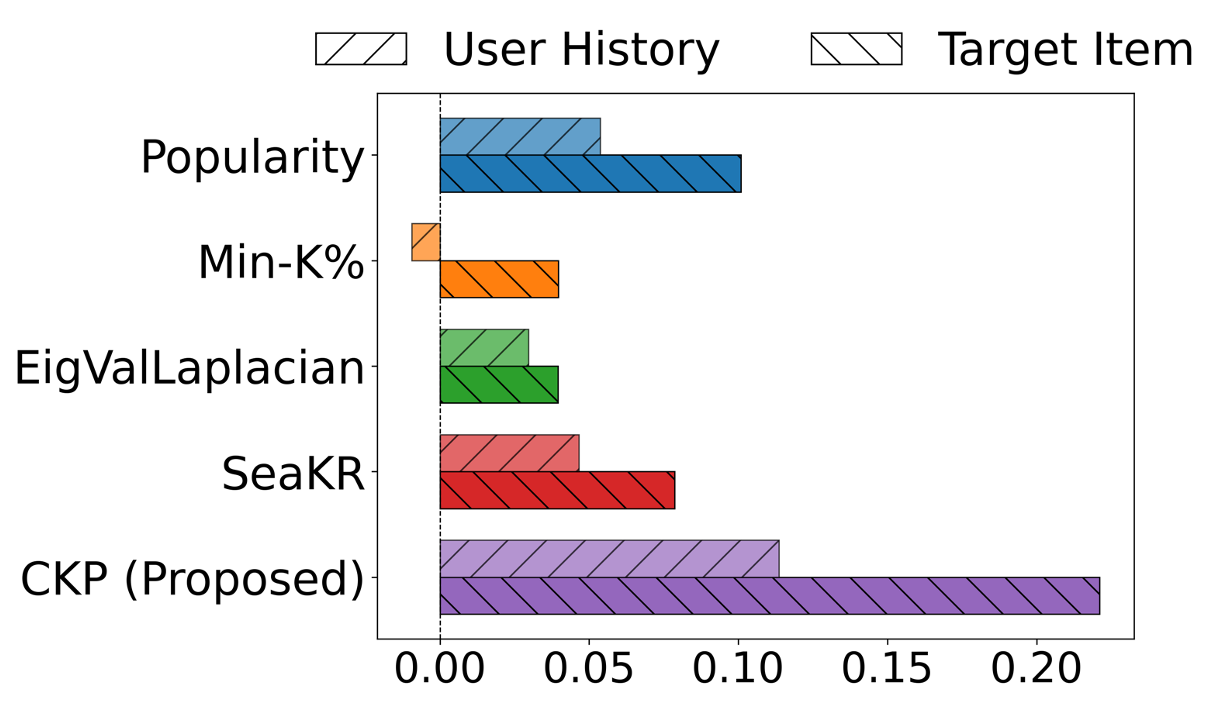}
    \Description{}
    \vspace{-4mm}
    \caption*{(a) Spearman correlation}
    \label{fig:beauty-corr}
\end{minipage}
\hspace{0.015\linewidth}
\begin{minipage}{0.64\linewidth}
    \centering
    \includegraphics[width=\linewidth]{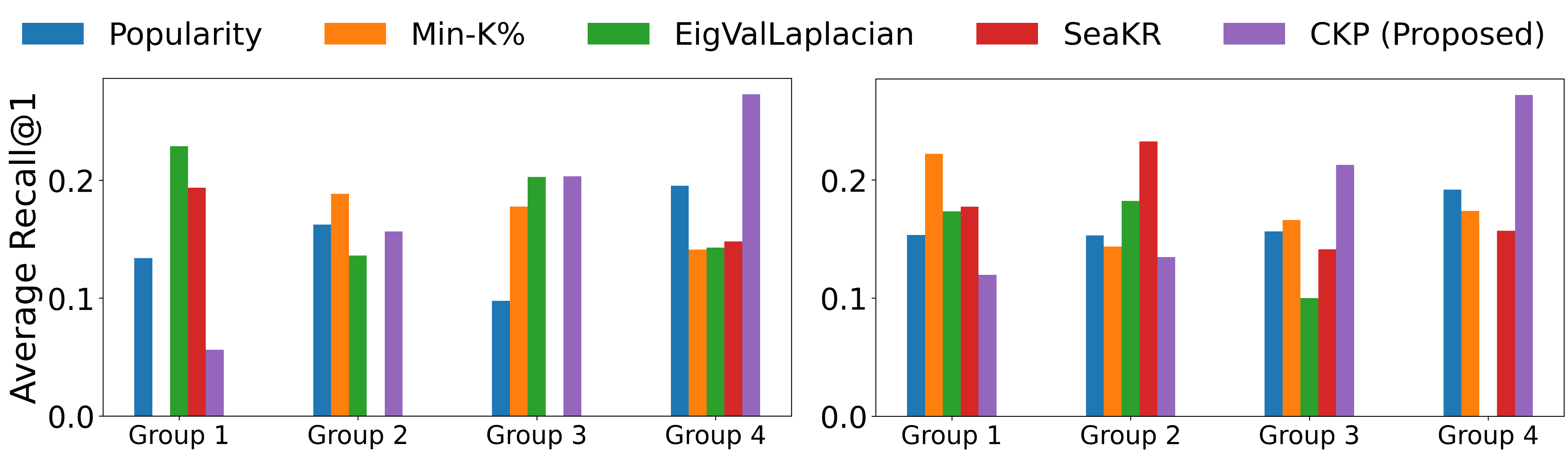}
    \Description{}
    \vspace{-4mm}
    \caption*{(b) Performance Alignment across Knowledge Levels}
    \label{fig:item-group}
\end{minipage}
% \vspace{-0.5mm}
\caption{Alignment between knowledge scores and recommendation quality on the A-Beauty dataset. (a) Spearman correlation between Recall@1 and knowledge scores obtained from five methods. (b) Average Recall@1 across four quantile bins grouped by knowledge scores at the item (left) and user (right) levels. 
Group 1 represents the lowest knowledge score bin, and Group 4 is the highest.
The proposed knowledge scoring method consistently aligns with recommendation performance across all perspectives.}
\label{fig:empirical}
\end{figure*}

\section{Analysis: Knowledge Scoring for LLM-based Recommendation}
Before presenting our framework, we empirically investigate the nature of the knowledge gap in LLM-based recommendation. 
Specifically, we address two fundamental questions: 
(1) Does simply injecting external information for all items resolve the knowledge deficit? 
(2) If selective augmentation is needed, can we rely on existing knowledge proxies to diagnose what the model knows?

\subsection{The Paradox of Uniform Augmentation} 
\label{sec:paradox} 
We examine the impact of uniform augmentation, where full item attributes are indiscriminately added for every item in the input prompt.
Figure~\ref{fig:ff-gain} illustrates the performance difference between uniform augmentation and no augmentation (i.e., each item is represented by its title, the simplest input form) across four datasets. 
Despite the richer input, we observe notable performance drops on three datasets, suggesting that uniform augmentation is not universally beneficial and may even degrade recommendation accuracy. 
Since uniform augmentation indiscriminately appends information even for items the model already knows, it leads to two critical issues. 
First, LLMs face known difficulties in interpreting information within extended contexts~\cite{lostinthemiddle, LLMRank-ECIR24}.
Consequently, augmented knowledge is often overlooked, making it difficult for the model to effectively utilize the provided information even for less-known items.
Second, the excessive prompt length substantially increases inference costs (e.g., API fees) and latency.
These observations underscore the necessity of selective augmentation: injecting information only where the model's internal knowledge is insufficient.

\subsection{The Misalignment of Existing Proxies} 
\label{sec:misalignment} 
To implement selective augmentation, it is essential to estimate the degree of knowledge that an LLM possesses for each item, a process we refer to as \textbf{knowledge scoring}.
We begin by evaluating whether existing proxies, ranging from simple heuristics to advanced adaptive RAG metrics, can serve as reliable indicators for this purpose.

\subsubsection{\textbf{Existing Proxies}}  
\label{sec:naive_proxy}
We consider four categories of proxies:
\begin{itemize}[leftmargin=*]
% \vspace{-\topsep} 
    \item \textbf{Popularity} refers to the item frequency in the interaction dataset. 
    This is a common heuristic for item exposure, similar to Wikipedia pageviews used in general domains~\cite{gap-ACL23, KRagRec25}.
    \item \textbf{Pre-training Data Detection} estimates whether an item was present in the LLM's pre-training corpus \cite{PDD-USENIX21, DCPDD-EMNLP24, Mink-ICLR24}. 
    We adopt \textbf{Min-K\%} \cite{Mink-ICLR24}, which computes the average log-probability over the bottom $k\%$ of tokens in the generated item title, conditioned on the target domain (e.g., In the movie domain, the title is:).
    \item \textbf{Uncertainty} measures the model's internal confidence in LLM predictions \cite{UE_entropy20, SAR_ACL24, CCP_ACL24, EigValLaplacian, SurveyUE_ACL25}.
    We utilize \textbf{EigValLaplacian} \cite{EigValLaplacian}, which computes the sum of Laplacian eigenvalues from a weighted graph constructed based on the semantic similarity of multiple sampled responses (e.g., descriptions generated for the movie "Titanic").
    \item \textbf{Adaptive Retrieval Score} represents the self-awareness of the LLM regarding its need for external information in Adaptive RAG research \cite{FLARE_EMNLP23, DRAGIN_ACL24, AdaptiveRAG_NACCL24, SeaKR}.
    We employ the scoring strategy from \textbf{SeaKR} \cite{SeaKR}, which measures the consistency of internal states across multiple responses generated from a query requesting an item description, where a lower score reflects a higher need for retrieval.
\end{itemize}\vspace{-\topsep}

\subsubsection{\textbf{Empirical Observations}} 
\label{prelim_analysis}
To serve as a reliable indicator of LLM's competence for recommendation, the proxy should be closely correlated with the recommendation performance.
We report the result with \texttt{Qwen-2.5-7B} with no prompt augmentation (i.e., each item is represented by its title, the simplest input form).
Similar tendencies are observed with other models.
Our findings reveal two key observations:

\vspace{2mm} \noindent 
% \vspace{0.02in} \noindent
\textbf{O1. Existing proxies correlate weakly with ranking performance.} 
We first evaluate whether the proxies align with actual recommendation quality by computing the Spearman correlation between each proxy score and the model’s recall (Recall@1).
For a comprehensive analysis, we analyze the correlations at both item- and user-levels.\footnote{For item-level, we compute the correlation across all test items. For user-level, we compute the average proxy score over the items in each user’s history and correlate it across all users.}
The results are presented in Figure~\ref{fig:empirical}(a).
Interestingly, we observe that the simplest heuristic, popularity, exhibits the highest correlation, whereas sophisticated, recent techniques yield weak correlations.

\begin{itemize}[leftmargin=*]
% \vspace{-\topsep} 
    \item \textbf{Popularity:} 
    As a heuristic derived solely from dataset statistics, popularity acts as an external signal that does not necessarily align with the LLM's actual exposure during pretraining. 
    Therefore, it cannot capture the full complexity of the model's internal knowledge, underscoring the need for a model-centric estimation that directly reflects the LLM's prediction behaviors.
    
    \item \textbf{Min-K\%:} 
    Since this method relies on token-level likelihoods of item titles, it is highly susceptible to surface-level biases.
    Item titles are typically short and lack sufficient context \cite{zero-shot-rec-COLING'25}, which leads to two major biases: 
    (i) \textit{Lexical Bias}, where high likelihood may merely reflect familiarity with common words (e.g., "New York") rather than genuine understanding of the item, and (ii) \textit{Length Bias}, where longer titles tend to receive lower total likelihoods due to the greater accumulation of negative log-probabilities, thereby penalizing those items regardless of the model's knowledge.

    \item \textbf{EigValLaplacian and SeaKR:} 
    Since these methods rely on generating item descriptions, they are inherently sensitive to prompt design. 
    Furthermore, the excessive length of generated descriptions incurs prohibitive computational costs, limiting scalability across large item catalogs. 
    Most critically, LLMs frequently exhibit unwarranted confidence by producing fluent descriptions even for completely unknown concepts~\cite{KUQ}, which yields unreliable signals as the model fails to distinguish between genuine knowledge and hallucinated familiarity.

\end{itemize}
% \vspace{-\topsep}
\noindent
Critically, all of these proxies focus on measuring item-specific knowledge in isolation. 
Consequently, this approach fails to capture \textit{collaborative patterns} such as co-consumption behaviors that are critical for recommendation.
Furthermore, given that the target domain knowledge represents a tiny fraction of the LLM's vast open-world knowledge, querying it without sufficient context often fails to \textit{activate the relevant information}.

\vspace{2mm} \noindent 
% \vspace{0.02in} \noindent
\textbf{O2. Existing proxies fail to reflect group-level differences in model competence.}
For a more in-depth understanding, we group target items and users into four quantile bins based on their knowledge scores and compute the average Recall@1 within each group.
A reliable proxy should exhibit a monotonic increase in performance across groups (i.e., higher knowledge scores $\rightarrow$ higher recall).
As shown in Figure~\ref{fig:empirical}(b), the existing proxies yield inconsistent performance across bins, failing to reveal a meaningful relationship between knowledge and performance.
In contrast, our proposed method (CKP) demonstrates a clear monotonic increase, confirming its effectiveness as a reliable knowledge indicator.

\vspace{2mm} \noindent
\textbf{Summary.} Our analysis shows that (1) uniform augmentation is suboptimal for mitigating the knowledge gap and improving recommendation quality, and (2) existing proxies used in other domains are not suitable for knowledge scoring in recommendation tasks.

\section{Methodology}
We introduce \proposed, a selective augmentation framework with comparative knowledge probing.
It consists of two main stages: (1) \textbf{knowledge scoring} that estimates the LLM's knowledge on each item for recommendation (\cref{method:ke}), and (2) \textbf{selective augmentation} that enriches the knowledge for lesser-known items (\cref{method:sa}).

\subsection{\textbf{Knowledge Scoring for Recommendation}}
\label{method:ke}
Our goal is to develop a knowledge scoring strategy tailored to LLM-based recommendations.
We adopt a likelihood-based approach, inferring the LLM's knowledge based on the likelihood it assigns to an item given interaction contexts.

\subsubsection{\textbf{Overview}}
For each item $t$, we construct its \textit{interaction context windows} $\mathcal{W}_t$ from interaction histories $\mathcal{H}$.
Each window $w \in \mathcal{W}_t$ is a sequence of items preceding $t$, providing its contexts.
Ideally, the knowledge score should be aggregated over the entire set $\mathcal{W}_t$ to fully capture the item's contextual patterns.
However, as this entails excessive computational costs, we employ a \textit{popularity-stratified sampling} strategy to obtain a representative subset $\widetilde{\mathcal{W}}_t \subset \mathcal{W}_t$. 
Specifically, we partition $\mathcal{W}_t$ into three quantile bins based on the average popularity of items in each window, and then uniformly sample from each bin.
This strategy ensures computational efficiency while preserving a balanced coverage of diverse interaction patterns.
For each window, we build a prompt $\mathcal{P}_w$ and measure the LLM's generation preference as a conditional probability, $P_{\text{LLM}}(t \mid \mathcal{P}_w)$.
The item-level knowledge score $K(t)$ is defined as the average of these probabilities over all sampled windows:
\begingroup
\begin{equation}
K(t) := \frac{1}{|\widetilde{\mathcal{W}_t}|} \sum_{w \in \widetilde{\mathcal{W}_t}} P_{\text{LLM}}(t \mid \mathcal{P}_w)
\label{eq:aggr_ke}
\end{equation}
\endgroup
The key challenges lie in the design of: (i) the prompt $\mathcal{P}_w$, including both instruction formulation and output structure, and (ii) the likelihood function $P_{\text{LLM}}(t \mid \mathcal{P}_w)$, particularly how to extract a reliable knowledge proxy from the generated output.
Our design is guided by three desiderata for recommendation-tailored scoring:
% \vspace{-\topsep}
\begin{itemize}[leftmargin=*] 
    \item[(1)] \textbf{Interaction-based contextualization:} 
    The scoring should reflect the LLM's behavior conditioned on specific interaction contexts, rather than on the coarse-grained conditions (e.g., domain).

    \item[(2)] \textbf{Ranking-oriented scoring:} As recommendation is inherently a ranking task that orders probable items, the scoring should capture the LLM’s ability to assign appropriate ranks.

    \item[(3)] \textbf{Robustness to surface-level bias:} The scoring should be robust against the inherent generation biases of LLMs (e.g., lexical preference, output length), as discussed in 
    \cref{prelim_analysis}.
\end{itemize}
% \vspace{-\topsep}
Guided by these desiderata, we introduce two knowledge scoring methods: a baseline Direct Knowledge Probe (DKP), and a more advanced Comparative Knowledge Probe (CKP).

\subsubsection{\textbf{Naive Approach: Direct Knowledge Probe (DKP)}}
As the simplest instantiation, DKP estimates the generation probability of the item title, given its interaction window.
The prompt $\mathcal{P}^{\text{DKP}}_w$ includes a window~$w$ and an instruction for next item prediction, e.g., ``Given $\{w\}$, the next item is:''.
The likelihood is obtained by aggregating the token probabilities of the item title $y = (y_1, \dots, y_L)$:
\begingroup
\begin{equation}
P_{\text{LLM}}(t \mid \mathcal{P}^{\text{DKP}}_w) = \exp \left( \sum_{j=1}^{L} \log P(y_j \mid \mathcal{P}^{\text{DKP}}_w, y_{<j}) \right)
\label{eq:dkp_prob}
\end{equation}
\endgroup
Substituting $P_{\text{LLM}}$ into Eq.~\ref{eq:aggr_ke} yields the item-level knowledge score for DKP.
While DKP satisfies the first desideratum by leveraging user interaction context, it fails the others: it considers each item separately, making it less aligned with the ranking-oriented nature of recommendation.
Also, it remains susceptible to surface-level biases, which can inflate scores for item titles containing verbose or frequently occurring tokens.
To overcome these limitations, we introduce our proposed method, CKP.

\subsubsection{\textbf{Comparative Knowledge Probe (CKP)}}
Our key idea to meet the remaining two desiderata is to reframe the scoring task as a \textit{relative comparison problem} based on \textit{content-neutral identifiers}.

\smallsection{Fine-grained comparison set.}
For each item $t$, we construct a \textit{comparison set} $C(t)$, consisting of $t$ and several distractor items.
The LLM will be instructed to rank this set given the interaction window~$w$.
Compared to querying each item independently, this approach is better aligned with the ranking nature of recommendation.
To achieve this, we employ a hybrid sampling strategy that combines random and semantic distractors.

First, we include $n$ \textit{random distractors} ($\mathcal{C}_{\mathrm{rand}}$), sampled uniformly from the item set $\mathcal{I}$.
These serve as \textbf{easy cases}, providing item diversity to ensure the model maintains the capability to discriminate the target from irrelevant noise.
\begingroup
\begin{equation}
\mathcal{C}_{\mathrm{rand}}(t) \sim
    \text{UniformSample}\bigl(\mathcal{I}\setminus(\{t\} \cup w),\, n\bigr)
\label{eq:rand_neg}
\end{equation}
\endgroup
Second, we include $m$ \textit{semantic distractors} ($\mathcal{C}_{\mathrm{sem}}$)—items that are semantically similar to $t$ but not valid for the current context.
These serve as \textbf{hard negatives} designed to mitigate the model's tendency to overestimate its knowledge.
By providing these \textbf{plausible alternatives}, we force the model to demonstrate fine-grained reasoning: it allows the model to concentrate probability on the target \textit{only when} it possesses sufficient discriminative knowledge to reject these semantically similar but contextually incorrect items.
They are selected based on the cosine similarity of text embeddings $\mathbf{e}$, derived from item titles and attributes:\footnote{As the simplest choice, we use Sentence-BERT \cite{reimers2019sentence}.} 
\begingroup
\begin{equation}
\mathcal{C}_{\mathrm{sem}}(t) = \operatorname{Top}-m_{j \in \mathcal{I} \setminus (\{t\} \cup w)} \left( \cos(\mathbf{e}_t, \mathbf{e}_j) \right)
\label{eq:hard_neg}
\end{equation}
\endgroup
The final set is $C(t) = \{t\} \cup \mathcal{C}_{\mathrm{rand}}(t) \cup \mathcal{C}_{\mathrm{sem}}(t)$.
This hybrid design balances item diversity and ranking difficulty, providing comparison sets that are comprehensive and challenging.

\smallsection{Identifier-based top‑1 estimation.}
Given the comparison set, a standard scoring method is to instruct the LLM to generate a ranked list of item titles provided in the prompt.
The probability of placing the target item $t$ (i.e., the true next item) near the top of the list can then serve as its knowledge score.
However, such an approach has two limitations.
First, as discussed earlier, it is susceptible to surface-level biases arising from the token composition of item titles.
Second, the ranking process itself suffers from a structural flaw due to the autoregressive nature of generation: items ranked later in the list are selected from a smaller remaining pool, leading to inflated probabilities simply because fewer~alternatives~remain.\footnote{For instance, in a set of three items, the probability of the last-ranked item is calculated from a pool of only one remaining option, artificially inflating its score}

We propose a simple yet effective solution by re-designing both the prompting scheme—the instruction and output structure—and the likelihood function.
Specifically, we randomly shuffled the elements in $C(t)$\footnote{This randomization mitigates positional bias~\cite{LLMRank-ECIR24}, ensuring that the model's selection is driven by content rather than the order of presentation.} and assign content-neutral identifiers (e.g., [A], [B]) to each item.
These identifiers are decoupled from the original item titles, effectively removing surface-level biases while remaining computationally efficient.
Then, we instruct the LLM to pinpoint the \textit{single most preferred item} from $C(t)$, instead of generating a full ranking.
The likelihood is computed based on the top-1 selection probability, thereby avoiding the spurious artifacts of full ranking.

% the listwise probability model
The prompt $\mathcal{P}^{CKP}_{w}$ includes a window $w$, the comparison set $C(t)$, and an instruction such as ''Choose a single identifier of the most preferred item''.
The LLM output for this prompt would be an index (e.g.,[A]) for an item in $C(t)$.
Then, with the output logit values $z_i \in \mathbb{R}$ for each $i \in C(t)$, we define the top-1 selection likelihood from the $C(t)$ based on the list-wise ranking model~\cite{cao2007learning}:
\begingroup
\setlength{\abovedisplayskip}{3pt}
\setlength{\belowdisplayskip}{3pt}
\begin{equation}
P_{\text{LLM}}(t \mid \mathcal{P}^{CKP}_{w}) = \frac{\phi(z_t)}{\phi(z_t) + \sum_{j \in C(t)\setminus \{t\} } \phi(z_j)}
\label{eq:ckp_prob_v2}
\end{equation}
\endgroup
where $\phi(\cdot)$ is an increasing function, we adopt the exponential function.

CKP effectively measures recommendation-tailored knowledge by explicitly prompting for top-1 selection.
It also satisfies all three desiderata: conditioning on $w$ for contextualization, selecting the top-1 from $C(t)$ for ranking alignment, and using context-neutral identifiers for robustness against surface-level biases.
Replacing $P_{\text{LLM}}$ in Eq.\ref{eq:aggr_ke} derives the knowledge score of CKP.

\begin{table}[t]
% \captionsetup{skip=3pt}
\caption{An example prompt used for calculating knowledge scores. All prompts are prefixed with a domain-specific system instruction (e.g., "You are a helpful assistant for \texttt{[DOMAIN]} recommendations.").}
\label{tab:prompts}
\centering
\small
\renewcommand{\arraystretch}{1.3}
\resizebox{\linewidth}{!}{%
\begin{tabular}{l|p{0.85\linewidth}}
\toprule
\textbf{Metric} & \textbf{Input Template} \\
\midrule
\textbf{DKP} & 
\texttt{The user's interaction history is as follows: [HISTORY (Item Titles]} \newline
\texttt{The next item is: [TARGET (Item Title)]} \\
\midrule
\textbf{CKP} & 
\texttt{Your task is to recommend the top-1 item from the candidate set based on the user's purchase history.} \newline
\texttt{You must only respond with the single identifier of the recommended item.} \newline
\texttt{PURCHASED ITEMS: [HISTORY (Item Titles)]} \newline
\texttt{CANDIDATE ITEMS: [COMPARISON SET (ID, Item Titles)]} \newline
\texttt{Candidate [} \\
\bottomrule
\end{tabular}%
}
% \vspace{-0.3cm}
\end{table}
\subsection{\textbf{Knowledge-aware Selective Augmentation}}
\label{method:sa}
We now enhance the LLM-based recommenders via selective knowledge augmentation.
Given the task setup (\cref{subsec:definition}), our focus lies in answering two key questions: \textit{what to augment} (\cref{subsub:APS}) and \textit{with which information} (\cref{subsub:RMS}), to maximize recommendation accuracy.
The final prompting process is provided in \cref{subsub:final_prompt}.

\subsubsection{\textbf{What to augment: Augmentation Priority Score}}
\label{subsub:APS}
While our knowledge score effectively reveals the LLM's competence, we note that other factors should also be considered when prioritizing which items to augment.
In recommendation tasks, it is well known that an item’s importance varies with factors such as its recency in the user history and its popularity~\cite{petrov2024rss, abbattista2024enhancing, SASRec};
lacking knowledge about such items can have a greater negative impact on recommendation accuracy.
By consolidating these factors, we define the \textit{Augmentation Priority Score (APS)} for each item $i$ as:
\begingroup
\begin{equation}
\mathrm{APS}(i) = (1-k(i)) \cdot f(i) \cdot r(i)
\label{eq:aps}
\end{equation}
\endgroup
This score consists of three components.\footnote{Each component is log-transformed, and min-max normalized to match the scale.}
The first term $(1-k(i))$ represents the LLM’s knowledge deficiency, where $k(i)$ is the normalized value of $K(i)$.
The second term $f(i)$ represents the interaction frequency, prioritizing statistically prominent items.
Finally, the recency score \(r(i) = \exp(-\lambda \cdot \mathrm{pos}(i))\) assigns higher weights to more recent interactions, where $pos(i)$ is the item's reverse chronological position (i.e., the most recent item has $pos(i)=0$) and $\lambda$ is a decay parameter.
This APS will be used to decide items to be augmented, which will be further explained in \cref{subsub:final_prompt}.

\subsubsection{\textbf{With which information: Reference Matching Score}}
\label{subsub:RMS}
After deciding the augmentation targets, we need to enrich each with information most beneficial for recommendation.
A natural starting point is the textual attributes of each item in $\mathcal{T}$.
Moreover, we note that items can be better understood in relation to others.
Specifically, semantically or behaviorally related items may provide valuable signals that complement the target item's standalone information.
We refer to such items as \textit{reference items} and incorporate their information for augmentation as well.

The reference items are obtained via \textit{Reference Matching Score (RMS)}, which quantifies the suitability of each item \( r \in \mathcal{I} \setminus (\{t\} \cup {H}^u \cup {C}^u ) \) as a knowledge-supporting reference for a target item \( t \):
\begingroup
\begin{equation}
\mathrm{RMS}(t, r) = k(r) \cdot s(t, r) \cdot c(t, r)
\label{eq:rms}
\end{equation}
\endgroup
This score also consists of three components, each properly normalized.
First, the normalized knowledge score $k(r)$ prioritizes items that are well understood by the LLM.
Second, the semantic similarity \( s(t, r) \) is measured as the cosine similarity between their text embeddings, as used in Eq. \ref{eq:hard_neg}.
Third, the co-consumption score \( c(t, r) \) captures behavioral proximity, computed as the co-occurrence frequency in the interaction histories.\footnote{\( c(t,r) = \frac{\text{freq}(t, r)}{\sqrt{\text{freq}(t)} \cdot \sqrt{\text{freq}(r)}}\), where \text{freq}(t,r) is the co-occurrence count within a sliding window of size 2, and \text{freq}(i) is the total number of windows containing item $i$.}
Our RMS design ensures that the reference items are semantically and behaviorally related to the target item, while also being well understood by the LLM.

Note that we do not augment the full details of these reference items.
Instead, we guide the LLM by providing minimal cues through their titles.
Since the LLM already possesses sufficient knowledge about them, this can activate the relevant information from its parameters, allowing the LLM to naturally reference it without consuming much input context budget.

\noindent
\textbf{Context-aware Variant.}
We acknowledge that the proposed RMS is inherently \textit{context-agnostic}; it produces the same set of references for a given target item $t$, regardless of the user's specific history.
However, since items possess multifaceted attributes, the ideal reference may vary depending on the user's specific context.
To address this, we explore a context-aware variant that personalizes the reference selection process by incorporating the user's sequential interaction patterns.
Specifically, we modulate the RMS score using the user-item similarity from the sequential retriever (e.g., SASRec~\cite{SASRec}) already employed for candidate generation.
This approach leverages existing latent representations without requiring a separate encoder.
Formally, the score is adjusted as $\mathrm{RMS}_{\text{ctx}} = \mathrm{RMS}(t, r) \cdot \cos(\mathbf{h}^u, \mathbf{e}^r)$, where $\mathbf{h}^u$ and $\mathbf{e}^r$ are the embeddings from the retrieval model.

\begin{algorithm}[t]
\caption{Selective Knowledge Augmentation}
\label{alg:augmentation}
\KwIn{
Original prompt with $H^u$ and $C^u$, item textual attributes $\{a^i\}$, normalized knowledge scores $k(i)$}
\KwOut{Augmented prompt}
% \vspace{0.3em}

Compute $\mathrm{APS}(i)$ for each item $i \in H^u \cup C^u$ \,\,(Eq.\ref{eq:aps})

Select top-$K_{\text{aug}}$ items with highest APS as targets $\mathcal{I}_{aug}$

\ForEach{$t \in \mathcal{I}_{aug}$}{
    Add textual attributes $\{a^t\}$ to the prompt
    
    Identify top-$K_{\text{ref}}$ reference items using RMS$(t, r)$ \,\,(Eq.\ref{eq:rms})
    
    Add titles of reference items to the prompt
}
\end{algorithm}

\subsubsection{\textbf{Final Prompt Construction}}
\label{subsub:final_prompt}
The detailed augmentation process is presented in Algorithm~\ref{alg:augmentation}.
The number of augmentation targets ($K_{\text{aug}}$) is empirically set, considering resource constraints such as GPU memory and the LLM's context budget.
For reference items, we find that a small number ($K_{\text{ref}} \leq 3$) is typically sufficient.
A conceptual example of the augmentation is provided below:

\begin{table}[h]
\small
    \centering
    \resizebox{1.0\linewidth}{!}{
    \begin{tabular}{|C|}
    \hline
    \textbf{USER HISTORY}:
    ["The Witcher 3", "Elden Ring", ..., "Salt and Sanctuary"] \\[0.1em]

    \textbf{CANDIDATE ITEMS}:
    ["Dark Souls", ..., "Hollow Knight"] \\ \hline
    
    \textbf{AUGMENTATION TARGET}: "Salt and Sanctuary"\\[0.1em]

    \textbf{AUGMENTED INFORMATION}: \\
    - \textit{Textual Attributes}: "A 2D action role-playing game that combines fast, brutal combat with richly developed RPG mechanics." \\[0.3em]
    - \textit{Reference Items}: ["Blasphemous", "Dead Cells"]
\\ \hline 
    \end{tabular}}
    % \vspace{-0.3cm}
\end{table}

\noindent

In sum, based on the knowledge scores, \proposed selectively injects additional information—item attributes and related reference items—where it is most needed.
This enables the LLM to utilize its context budget more effectively by allocating more capacity to those that benefit most from knowledge supplementation.

\vspace{2mm}
\noindent
\textbf{Remarks: inference efficiency of \proposed.}
Although \proposed incorporates multiple factors such as knowledge scores and co-consumption score, these values are \textit{precomputed} and \textit{stored offline}. 
Thus, they can be accessed via simple lookup operations without introducing meaningful runtime overhead.
In practice, \proposed incurs negligible additional inference latency. 
Notably, compared to uniform augmentation, it \textbf{reduces the total inference latency} by avoiding unnecessary input tokens (\cref{subsub:eff}).

\begin{table*}[t!]
\centering
\begingroup
\small
% \setlength{\tabcolsep}{5.0pt}
% \captionsetup{skip=3pt}
\caption{Overall performance (Recall@1). Each subcolumn shows results with Random (R) and SASRec (S) candidates. \textcolor{red}{Red} color indicates no improvement over No Augment.
The best and second-best results are marked in \textbf{bold} and \underline{underlined}, respectively.}
\label{tab:overall_performance}
\renewcommand\theadfont{\bfseries}
\resizebox{\linewidth}{!}{%
\begin{tabular}{ll *{10}{c@{\hspace{4pt}}c} c@{\hspace{4pt}}c}
\toprule
\textbf{Dataset} & \textbf{LLM} & 
\multicolumn{2}{c}{\thead{No Augment}} & 
\multicolumn{2}{c}{\thead{Uniform-Meta}} & 
\multicolumn{2}{c}{\thead{Uniform-Wiki}} & 
\multicolumn{2}{c}{\thead{Selective\textsubscript{Acc}}} & 
\multicolumn{2}{c}{\thead{Selective\textsubscript{Pop}}} & 
\multicolumn{2}{c}{\thead{Selective\textsubscript{MinK}}} &
\multicolumn{2}{c}{\thead{Selective\textsubscript{EigV}}} &
\multicolumn{2}{c}{\thead{Selective\textsubscript{SeaKR}}} &
\multicolumn{2}{c}{\thead{Selective\textsubscript{Self}}} &
% \multicolumn{2}{c}{\thead{\proposedtwo}} & 
\multicolumn{2}{c}{\thead{\proposed}} &
\multicolumn{2}{c}{\thead{Improv. (\%)}} \\
\cmidrule(lr){3-4} \cmidrule(lr){5-6} \cmidrule(lr){7-8} \cmidrule(lr){9-10} \cmidrule(lr){11-12} \cmidrule(lr){13-14} \cmidrule(lr){15-16} \cmidrule(lr){17-18} \cmidrule(lr){19-20} \cmidrule(lr){21-22} \cmidrule(lr){23-24} 
 & & R & S & R & S & R & S & R & S & R & S & R & S & R & S & R & S & R & S & R & S & R & S \\
\midrule
\multirow{4}{*}{\textbf{A-Beauty}} 
& Llama-8B   
& 0.141 & 0.075 & 0.181 & \textcolor{red}{0.070} & \textcolor{red}{0.093} & \textcolor{red}{0.073} & 0.174 & \underline{0.221} & 0.161 & 0.212 & \underline{0.191} & 0.199 & 0.182 & 0.207 & 0.166 & 0.191 & \textcolor{red}{0.098} & 0.110 & \textbf{0.233} & \textbf{0.246} & 22.0 & 11.3 \\
& Mistral-7B 
& 0.098 & 0.084 & \textcolor{red}{0.074} & \textcolor{red}{0.069} & \textcolor{red}{0.074} & \textcolor{red}{0.068} & \textcolor{red}{0.065} & \textcolor{red}{0.074} & \textcolor{red}{0.072} & 0.097 & \textcolor{red}{0.075} & \underline{0.100} & \underline{0.102} & \textcolor{red}{0.079} & \textcolor{red}{0.060} & \textcolor{red}{0.071} & \textcolor{red}{0.044} & \textcolor{red}{0.038} & \textbf{0.113} & \textbf{0.119} & 10.8 & 19.0 \\
& Qwen-7B    
& 0.170 & 0.111 & 0.207 & \textcolor{red}{0.099} & 0.179 & 0.133 & 0.209 & 0.140 & \underline{0.216} & 0.138 & 0.199 & 0.141 & 0.206 & 0.146 & 0.201 & \underline{0.150} & \textcolor{red}{0.128} & \textcolor{red}{0.091} & \textbf{0.267} & \textbf{0.166} & 23.6 & 10.7 \\
& Qwen-32B  
& 0.295 & 0.218 & \textcolor{red}{0.293} & \textcolor{red}{0.213} & 0.348 & \underline{0.237} & 0.351 & 0.226 & 0.329 & 0.218 & \underline{0.361} & 0.221 & 0.349 & \textcolor{red}{0.217} & 0.339 & 0.226 & \textcolor{red}{0.271} & \textcolor{red}{0.216} & \textbf{0.407} & \textbf{0.257} & 12.7 & 8.4 \\
\midrule
\multirow{4}{*}{\textbf{A-Gift}} 
& Llama-8B   
& 0.062 & 0.044 & 0.063 & 0.048 & \textcolor{red}{0.050} & \textcolor{red}{0.040} & 0.156 & 0.085 & 0.129 & 0.085 & \underline{0.157} & \underline{0.100} & 0.145 & 0.078 & 0.147 & 0.077 & 0.074 & 0.046 & \textbf{0.167} & \textbf{0.110} & 6.4 & 10.0 \\
& Mistral-7B 
& 0.084 & 0.049 & \underline{0.101} & \textcolor{red}{0.037} & \textcolor{red}{0.058} & \underline{0.051} & 0.096 & \textcolor{red}{0.048} & \textcolor{red}{0.081} & \textcolor{red}{0.042} & \textcolor{red}{0.077} & \textcolor{red}{0.039} & \textcolor{red}{0.058} & 0.049 & \textcolor{red}{0.058} & 0.050 & \textcolor{red}{0.031} & \textcolor{red}{0.022} & \textbf{0.114} & \textbf{0.059} & 12.9 & 18.0 \\
& Qwen-7B    
& 0.095 & 0.056 & \textcolor{red}{0.077} & 0.058 & \textcolor{red}{0.090} & \textcolor{red}{0.051} & 0.096 & 0.063 & 0.098 & \textcolor{red}{0.054} & \textcolor{red}{0.088} & \underline{0.070} & \textcolor{red}{0.090} & \underline{0.070} & \textcolor{red}{0.091} & 0.069 & \underline{0.099} & 0.061 & \textbf{0.112} & \textbf{0.083} & 13.1 & 18.6 \\
& Qwen-32B  
& \underline{0.165} & 0.090 & \textcolor{red}{0.155} & \textcolor{red}{0.082} & \textcolor{red}{0.149} & \textcolor{red}{0.057} & \textcolor{red}{0.156} & \underline{0.106} & \textcolor{red}{0.153} & 0.102 & \textcolor{red}{0.155} & 0.104 & \textcolor{red}{0.157} & 0.094 & \textcolor{red}{0.153} & 0.103 & \textcolor{red}{0.138} & 0.093 & \textbf{0.175} & \textbf{0.119} & 6.1 & 14.4 \\
\midrule
\multirow{4}{*}{\textbf{ML-1M}} 
& Llama-8B   
& 0.133 & 0.067 & \textcolor{red}{0.104} & \textcolor{red}{0.051} & \textcolor{red}{0.049} & \textcolor{red}{0.055} & \textcolor{red}{0.125} & \textcolor{red}{0.058} & \underline{0.136} & \textcolor{red}{0.061} & \textcolor{red}{0.095} & \underline{0.071} & 0.134 & \textcolor{red}{0.053} & \textcolor{red}{0.105} & \textcolor{red}{0.049} & \textcolor{red}{0.110} & \textcolor{red}{0.053} & \textbf{0.152} & \textbf{0.088} & 11.8 & 23.9 \\
& Mistral-7B 
& 0.089 & 0.038 & 0.109 & \textcolor{red}{0.033} & \textcolor{red}{0.048} & \underline{0.053} & \textcolor{red}{0.047} & 0.043 & 0.101 & 0.045 & \underline{0.117} & 0.052 & 0.113 & 0.045 & 0.103 & 0.038 & \textcolor{red}{0.034} & \textcolor{red}{0.022} & \textbf{0.125} & \textbf{0.065} & 6.8 & 22.6 \\
& Qwen-7B    
& 0.139 & 0.039 & \textcolor{red}{0.132} & \underline{0.065} & \textcolor{red}{0.073} & \underline{0.065} & \underline{0.154} & \textcolor{red}{0.021} & 0.149 & 0.054 & 0.140 & 0.058 & \textcolor{red}{0.128} & \textcolor{red}{0.032} & 0.153 & \textcolor{red}{0.018} & 0.143 & \textcolor{red}{0.021} & \textbf{0.168} & \textbf{0.070} & 9.1 & 7.7 \\
& Qwen-32B  
& \underline{0.181} & 0.037 & \textcolor{red}{0.129} & \underline{0.049} & \textcolor{red}{0.125} & 0.040 & \textcolor{red}{0.164} & \textcolor{red}{0.029} & \textcolor{red}{0.172} & \textcolor{red}{0.031} & \textcolor{red}{0.163} & \textcolor{red}{0.031} & \textcolor{red}{0.169} & \textcolor{red}{0.035} & \textcolor{red}{0.167} & \textcolor{red}{0.031} & \textcolor{red}{0.173} & 0.043 & \textbf{0.201} & \textbf{0.054} & 11.0 & 10.2 \\
\midrule
\multirow{4}{*}{\textbf{Steam}} 
& Llama-8B   
& 0.078 & 0.042 & \textcolor{red}{0.063} & \textcolor{red}{0.041} & 0.086 & \textcolor{red}{0.039} & 0.111 & \textcolor{red}{0.029} & 0.122 & 0.054 & 0.095 & \underline{0.066} & 0.111 & \textcolor{red}{0.030} & \underline{0.141} & \textcolor{red}{0.032} & 0.097 & 0.048 & \textbf{0.152} & \textbf{0.072} & 7.8 & 9.1 \\
& Mistral-7B 
& 0.059 & 0.033 & \textcolor{red}{0.051} & \textcolor{red}{0.030} & \textcolor{red}{0.057} & 0.046 & \textcolor{red}{0.056} & 0.041 & 0.071 & \underline{0.055} & \textcolor{red}{0.057} & 0.044 & \underline{0.073} & \textcolor{red}{0.027} & 0.068 & \textcolor{red}{0.019} & \textcolor{red}{0.028} & 0.023 & \textbf{0.077} & \textbf{0.067} & 5.5 & 21.8 \\
& Qwen-7B    
& 0.108 & 0.043 & \textcolor{red}{0.059} & 0.050 & \textcolor{red}{0.061} & 0.046 & \underline{0.133} & \textcolor{red}{0.034} & 0.127 & 0.043 & 0.112 & \underline{0.057} & 0.119 & \textcolor{red}{0.035} & 0.119 & \textcolor{red}{0.030} & 0.119 & \textcolor{red}{0.028} & \textbf{0.148} & \textbf{0.063} & 11.3 & 10.5 \\
& Qwen-32B  
& 0.121 & 0.035 & \textcolor{red}{0.073} & 0.044 & \textcolor{red}{0.114} & 0.044 & \textcolor{red}{0.117} & 0.045 & 0.123 & 0.042 & \underline{0.126} & 0.043 & \textcolor{red}{0.120} & \underline{0.047} & \textcolor{red}{0.117} & 0.042 & \textcolor{red}{0.101} & 0.045 & \textbf{0.150} & \textbf{0.053} & 19.0 & 12.8 \\
\bottomrule
\end{tabular}
}
\endgroup
% \vspace{-0.2cm}
\end{table*}

% \textbf{Dataset} & \textbf{LLM} & 
% \multicolumn{2}{c}{\thead{LLMRank\textsubscript{T}}} & 
% \multicolumn{2}{c}{\thead{LLMRank\textsubscript{F}}} & 
% \multicolumn{2}{c}{\thead{Select\textsubscript{SASRec}}} & 
% \multicolumn{2}{c}{\thead{Select\textsubscript{Pop}}} & 
% \multicolumn{2}{c}{\thead{Select\textsubscript{Gen}}} &
% \multicolumn{2}{c}{\thead{SA\textsubscript{DKP}}} & 
% \multicolumn{2}{c}{\thead{SA\textsubscript{CKP}}} &
% \multicolumn{2}{c}{\thead{Improv. (\%)}} \\

\section{Experiments}
\subsection{Experimental Setup}

\begin{table}[hbt!]
\centering
% \captionsetup{skip=3pt}
\caption{Statistics of the datasets used in our experiments. Avg. Len denotes the average sequence length of users.}
\label{tab:data_statistics}
\resizebox{\linewidth}{!}{%
\begin{tabular}{lrrrrr}
\toprule
\textbf{Dataset} & \textbf{\#Users} & \textbf{\#Items} & \textbf{\#Inter.} & \textbf{Avg. Len} & \textbf{Attributes} \\
\midrule
\textbf{A-Beauty} & 4884 & 3948 & 16973 & 3.50 & \text{title, brand} \\
\textbf{A-Gift} & 3392 & 834 & 13503 & 4.01 & \text{title, brand, category}\\
\textbf{ML-1M} & 6040 & 3416 & 999611 & 161.85 & \text{title, genre}\\
\textbf{Steam} & 25859 & 4038 & 327097 & 12.93 & \text{title, genre, developer, specs}\\
\bottomrule
\end{tabular}}
\end{table}

\noindent
\textbf{Datasets.}
We use four public datasets: 
Amazon-Beauty (A-Beauty) \cite{ni2019justifying}, Amazon-Gift Cards (A-Gift)~\cite{ni2019justifying}, ML-1M~\cite{harper2015movielens}, and Steam~\cite{SASRec}.
For all datasets, we filter out users and items with fewer than five interactions, and items with missing textual features.
The interactions are sorted chronologically to form historical sequences.
The statistics for each preprocessed dataset are summarized in Table \ref{tab:data_statistics}.

\smallsection{Baselines.}
We compare various augmentation strategies, categorized into three groups.
\textbf{(a) No Augment} denotes the default setup, where each item is represented by its title.
\textbf{(b) Uniform Augmentation} indiscriminately enriches all items, the most widely used approach in the literature.
Specifically, we test two variants: \textbf{Uniform-Meta} (Attributes) and \textbf{Uniform-Wiki} (Wikipedia descriptions).
\textbf{(c) Selective Augmentation} prioritizes augmentation for certain items.
For fair comparison, all methods augment target items with item attributes and reference item titles.
We implement baselines corresponding to the proxies analyzed in \cref{sec:naive_proxy}:
\textbf{Selective\textsubscript{Pop}} (Popularity),
\textbf{Selective\textsubscript{MinK}} (Pre-training Detection via Min-K\%~\cite{Mink-ICLR24}),
\textbf{Selective\textsubscript{EigV}} (Uncertainty via EigValLaplacian~\cite{EigValLaplacian}), and
\textbf{Selective\textsubscript{SeaKR}} (Adaptive RAG via SeaKR~\cite{SeaKR}).
We also include \textbf{Selective\textsubscript{Acc}}, which uses Recall$@$1 from the candidate generation stage.\footnote{In the two-stage recommendation (\cref{subsec:definition}), the accuracy of the first-stage model can serve as a natural proxy. In this work, we use item-level Recall@1 of SASRec \cite{SASRec}.}
Additionally, we introduce \textbf{Selective\textsubscript{Self}}, a prompting-based baseline that operates via a two-stage inference without offline scoring, explicitly asking the LLM to identify unfamiliar items in the input prompt to apply selective augmentation.
Lastly, \textbf{\proposed} denotes the proposed frameworks, utilizing CKP.

\smallsection{Evaluation setup.}
We adopt the standard leave‑one‑out protocol \cite{LLMRank-ECIR24,SASRec}, where the last item in each user sequence serves as the ground‑truth item.
The evaluation task is to rank a candidate set of 20 items (1 ground-truth with 19 negatives) under two different settings:  
(i) \textbf{Random (R)}, with negatives from uniform sampling over unseen items~\cite{ALLMRec-KDD24, KRagRec25}, 
and (ii) \textbf{SASRec (S)}, with negatives from the top-19 predictions of a SASRec trained on each dataset~\cite{LLMRank-ECIR24}.
Following \cite{ALLMRec-KDD24}, we report Recall$@$1 of the ranking list from LLMs.

\smallsection{Implementation details.}
We evaluate our framework on Llama-3.1-8B~\cite{llama3}, Mistral-7B-v0.3~\cite{Mistral7}, Qwen2.5-7,32B~\cite{qwen2.5} and GPT-4o~\cite{gpt4o}.
We follow the prompting scheme from \cite{RLMRec-WWW24}.
An example prompt is provided in Table \ref{tab:prompt_structure}.
For fair comparison, we fix the history, candidates, and prompt template across all methods; only the content of the auxiliary information varies by augmentation strategy.

For each dataset, we randomly sample 1,500 users for testing. 
The remaining data is split into training and validation sets (9:1) for SASRec training, knowledge scoring, and hyperparameter tuning.
User histories are truncated to the recent 50 interactions.
For hyperparameters, we tune the model in certain ranges as follows:
(1) For SASRec, embedding dimension $\in \{64, 128\}$ and batch size $\in \{64, 128, 256\}$.
(2) For CKP, random ($n$) and semantic ($m$) distractors $\in \{0, 1, 2\}$.
(3) For APS, recency decay $\lambda \in \{0.7, 0.4, 0.1\}$ and augmentation targets $K_{\text{aug}} \in \{2, 3, 5, 10, 20, 40\}$.
(4) For RMS, reference items $K_\text{ref} \in \{1, 2, 3\}$.

\begin{table}[h]
% \captionsetup{skip=3pt}
\caption{Structure of the recommendation prompt.}
\label{tab:prompt_structure}
\small
    \centering
    \resizebox{1.0\linewidth}{!}{
    \begin{tabular}{|C|}
    \hline
    \textbf{INSTRUCTION}:
    % Your task is to recommend 20 games. \\[0.1em] \hline
    Your task is to recommend 20 games to a specific user from a candidate item set. \\[0.1em] \hline
    
    \textbf{PURCHASED ITEMS}:
    ["The Witcher 3", "Elden Ring", ..., "Salt and Sanctuary"] \\[0.1em]

    \textbf{CANDIDATE ITEMS}:
    ["A": "Dark Souls", ..., "T": "Hollow Knight"] 
    
    \textbf{AUXILIARY INFORMATION}: 
    [\{"title": "Salt and Sanctuary", "description": "A 2D action role-playing game ...", "title of similar game": "Blasphemous"\}] \\[0.1em] \hline

    \textbf{OUTPUT}:
    ["A", "C", ..., "T"] \\[0.1em] \hline 
    \end{tabular}}
    % \vspace{-0.3cm}
\end{table}

\subsection{Overall Performance}
Table~\ref{tab:overall_performance} shows performance across four datasets.
\proposed consistently outperforms all baselines across all datasets and LLMs.
First, uniform augmentation does not guarantee performance improvement.
They often fails to outperform the 'No Augment' (\textcolor{red}{red} in table), likely because indiscriminately adding all metadata can exceed the LLM's effective context budget, leading to information overload that obscures critical information.

Also, naive selective augmentation strategies relying on heuristics show inconsistent results across datasets and models, indicating that single-view proxies are not robust.
This again supports the necessity of a tailored augmentation strategy for recommendation.
Finally, \proposed achieves robust and superior performance in all settings.
By leveraging CKP for accurate knowledge estimation and guiding augmentation via APS and RMS, it provides targeted, effective enhancements beyond simple heuristics.

\begin{figure}[t]
\centering
% \vspace{-1mm}
\includegraphics[width=0.9\linewidth]{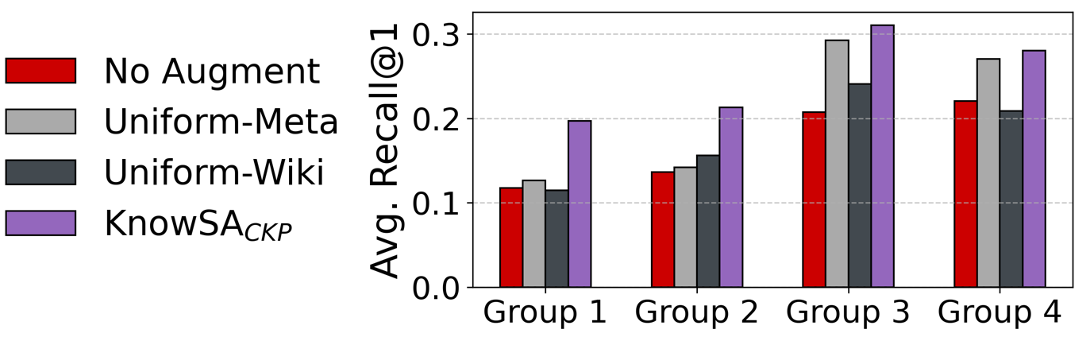}
\Description{}
% \vspace{-1.5mm}
\caption{Average Recall@1 across four quantile bins grouped by knowledge score (CKP) on A-Beauty dataset. Results of Qwen-7B with random candidates.}
\label{fig:group-recall-user}
% \vspace{-0.0cm}
\end{figure}

\noindent
\textbf{Performance by knowledge levels.}
We further analyze performance across users with varying levels of knowledge.
We group test users into four quantile groups based on the average CKP scores of items in the interaction history, with Group 1 representing the most knowledge-poor users.
Figure~\ref{fig:group-recall-user} shows that \proposed consistently achieves the highest accuracy across all groups.
Notably, the performance gap between \proposed and uniform augmentation is most pronounced in Groups 1–2.
This suggests that \proposed more effectively bridges the knowledge gap through selective augmentation guided by knowledge scores.
Further, it confirms that our targeted augmentation is highly effective in supplementing knowledge for users where LLMs have limited~competence.

\subsection{Study of \proposed}

\subsubsection{\textbf{Comparison with Frontier Models}}
\label{subsub:sa}
One might attribute the poor performance of the \textit{Selective\textsubscript{Self}} baseline to the limited reasoning capabilities of open-source models.
To investigate this, we conducted a comparative experiment using a frontier model, GPT-4o.
As shown in Table~\ref{tab:gpt_vs_qwen}, remarkably, even for GPT-4o, the self-asking strategy resulted in a performance drop compared to the \textit{No Augment} baseline, whereas Qwen-32B equipped with KnowSA\textsubscript{CKP} achieved the highest performance.
This suggests that accurately diagnosing knowledge gaps for numerous distinct items (up to 70 items, including 50 history and 20 candidate items) within a single prompt is an inherently difficult task, even for advanced LLMs.
These findings underscore the importance of a scoring strategy designed to diagnose knowledge gaps in the context of recommendation.

\begin{table}[t]
\centering
% \captionsetup{skip=0pt}
\caption{Performance comparison on A-Beauty dataset. We investigate whether a stronger LLM can effectively perform selective augmentation via self-asking. \textbf{Improv.} denotes the relative improvement in Recall compared to the No Augment baseline.}
\label{tab:gpt_vs_qwen}
\resizebox{0.95\columnwidth}{!}{%
\begin{tabular}{llccc}
\toprule
\textbf{Model} & \textbf{Method} & \textbf{R} & \textbf{S} & \textbf{Improv.} \\ \midrule
\multirow{3}{*}{\textbf{GPT-4o}} 
& No Augment & 0.316 & 0.265 & - \\
& Uniform-Meta & 0.349 & 0.269 & +10.4\% \\
& Selective\textsubscript{Self} & 0.300 & 0.251 & \textcolor{red}{-5.1\%} \\ \midrule
\multirow{3}{*}{\textbf{Qwen-32B}} 
& No Augment & 0.295 & 0.218 & - \\
& Uniform-Meta & 0.293 & 0.213 & -0.7\% \\
& \textbf{KnowSA\textsubscript{CKP} (Ours)} & \textbf{0.407} & \textbf{0.257} & \textcolor{blue}{\textbf{+38.0\%}} \\ \bottomrule
\end{tabular}%
}
% \vspace{-0.2cm}
\end{table}

\subsubsection{\textbf{Ablation Study}}
Table~\ref{tab:ablation} presents the results of various ablations.
First, we analyze CKP, our knowledge scoring strategy. 
Removing relative comparison, which replaces CKP with DKP, largely degrades performance, and removing semantic distractors leads to a slight drop—validating our comparative probing design.
Next, we assess our augmentation mechanisms.
For both APS and RMS, removing each proposed component results in performance degradation, with a particularly severe drop observed when the knowledge score is excluded, which supports the validity of~our~design.

The necessity of APS is particularly evident on ML-1M, a dataset with long historical interactions.
The severe performance drop of `w/o APS' highlights its critical role; prioritizing augmentations is essential in long contexts to ensure effective knowledge injection and to mitigate potential misinterpretations by LLMs~\cite{lostinthemiddle}.
For RMS, removing reference items entirely (`w/o RMS') degrades performance. 
Notably, co-consumption score proves to be vital, showing the importance of reflecting behavioral collaborative~patterns.

\begin{table}[t]
\centering
% \captionsetup{skip=3pt}
\caption{Ablation study on the components of the proposed method using Qwen-7B with random candidates.}
\label{tab:ablation}
\resizebox{1\linewidth}{!}{%
\begin{tabular}{lccc}
\toprule
\textbf{Method Variant} & \textbf{A-Beauty} & \textbf{A-Gift} & \textbf{ML-1M} \\
\midrule
\textbf{KnowSA\textsubscript{CKP} (ours)} & \textbf{0.267} & \textbf{0.112} & \textbf{0.168} \\
\midrule
\multicolumn{4}{l}{\textbf{Ablations on CKP}} \\
\quad \textit{w/o Relative Comparison} & 0.214 & 0.081 & 0.139 \\
\quad \textit{w/o Semantic Distractors}      & 0.219 & 0.110 & 0.151 \\
\addlinespace
\multicolumn{4}{l}{\textbf{Ablations on APS}} \\
\quad \textit{w/o APS \,\,(Uniform Aug. w/ ref. items)}& 0.194 & 0.072 & 0.087 \\
\quad \textit{w/o Knowledge Score}            & 0.213 & 0.095 & 0.151 \\
\quad \textit{w/o Interaction Frequency}      & 0.216 & 0.109 & 0.153 \\
\quad \textit{w/o Recency Score}              & 0.225 & 0.110 & 0.160 \\
\addlinespace
\multicolumn{4}{l}{\textbf{Ablations on RMS}} \\
\quad \textit{w/o RMS\,\,\,\,(Textual attributes only)}                        & 0.223 & 0.087 & 0.123 \\
\quad \textit{w/o RMS\,\,\,\,(Wikipedia description only)}                        & 0.079 & 0.032 & 0.063 \\
\quad \textit{w/o Knowledge Score}            & 0.188 & 0.095 & 0.146 \\
\quad \textit{w/o Semantic Similarity}        & 0.215 & 0.107 & 0.149 \\
\quad \textit{w/o Co-consumption Score}       & 0.169 & 0.109 & 0.135 \\
\quad \textit{w/o SASRec Contextualization} & 0.233 & 0.114 & 0.167 \\ 

\bottomrule
\end{tabular}}
% \vspace{-0.3cm}
\end{table}

\subsubsection{\textbf{Analysis on Long-tail Coverage and Bias}}
\begin{figure}[t]
\centering
% \vspace{-1mm}
\includegraphics[width=0.95\linewidth]{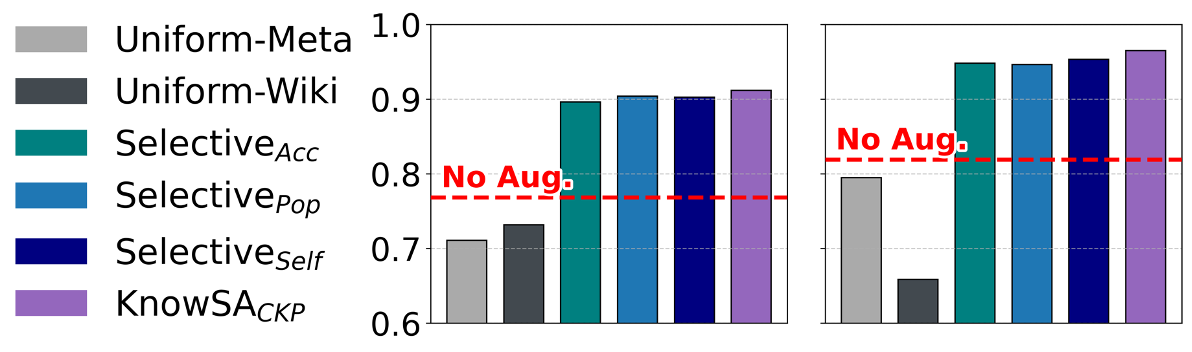}
\Description{}
% \vspace{-3mm}
\caption{Long-tail Coverage (LTC@10) on A-Beauty (left) and Steam (right), using Qwen-7B with random candidates.}
\label{fig:ltc-knowalign}
% \vspace{-0.2cm}
\end{figure}

To assess how our framework mitigates the knowledge gap for less-known items, we examine the coverage of long-tail items.
As a metric, we employ \textbf{Long-tail Coverage (LTC@K)}~\cite{LTC}, which measures the average fraction of long-tail items—defined as those in the bottom 80\% of popularity—appearing in a user's top-$K$ recommendation list.\footnote{$\text{LTC@K} = \frac{1}{|U_{\text{test}}|} \sum_{u \in U_{\text{test}}} |R^K_u \cap I_{LT}| / K$, where $I_{LT}$ is the long-tail item set, and $R_u^K$ is the top-$K$ list for user $u$. 
It follows the APLT metric~\cite{LTC}, adapted for top‑$K$ recommendation.
}
A higher LTC@K value indicates better coverage of long-tail items.
Figure~\ref{fig:ltc-knowalign} shows that selective augmentation generally improves long-tail coverage compared to uniform augmentation.
While \proposed yields a slight gain in LTC@K over other selective baselines, it achieves significantly higher recommendation accuracy in Table~\ref{tab:overall_performance}.
This shows that \proposed achieves a superior balance, enhancing recommendation diversity without sacrificing accuracy.
Furthermore, we investigate whether this improvement stems from an artificial bias, where the LLM over-recommends augmented items simply because they are enriched with extra knowledge.
We analyze the Top-1 prediction frequency of the \textit{augmentation targets} ($\mathcal{I}_{aug}$) within the candidate set.
Interestingly, compared to 'No Augment', \proposed reducing this frequency from 393 to 375 on A-Beauty using Qwen-7B.
This indicates that our augmentation provides clarifying knowledge, enhancing the model's ability to discern and reject inappropriate candidates, rather than simply boosting their ranks.

\subsubsection{\textbf{Efficiency Analysis}}
\label{subsub:eff}
We analyze the computational efficiency of \proposed from both offline preparation and online inference perspectives.
First, we address the scalability of the offline precomputation phase.
While calculating knowledge scores, CKP, for all item windows appears computationally intensive, our popularity-stratified sampling strategy (\cref{method:ke}) drastically reduces this cost.\footnote{For instance, on the ML-1M dataset, the number of sampled windows accounts for only 1\% of the total windows.}
As shown in Table~\ref{tab:scoring_cost}, our strategy consistently accelerates the scoring process across all datasets.
This computational efficiency ensures that precomputing knowledge scores is practically feasible, allowing them to be stored offline.
Second, regarding online inference, we evaluate the average number of input tokens per user history and the corresponding inference latency.
Since the augmentation factors are precomputed, they can be accessed with negligible runtime overhead.
Table~\ref{tab:efficiency} shows that the 'Uniform-Meta.' strategy incurs substantial overhead; indiscriminately adding all attributes significantly increases both token count and latency.\footnote{Note that latency does not increase linearly with input tokens; inference time is dominated by output decoding, while input processing is parallelized on GPUs~\cite{yang2024queueing}.}
In contrast, \proposed greatly reduces this overhead while still achieving the highest accuracy, as shown in Table~\ref{tab:overall_performance}.

\subsubsection{\textbf{Hyperparameter Analysis}}
\begin{figure}[t]
  \centering
  \begin{minipage}[t]{0.47\linewidth}
    \centering
    \includegraphics[width=0.8\linewidth]{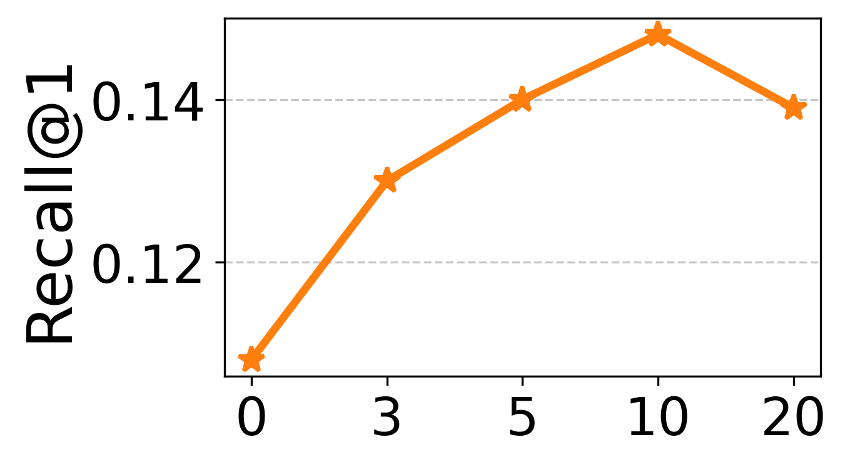}
    \Description{}
  \end{minipage}% \hfill
  \begin{minipage}[t]{0.47\linewidth}
    \centering
    \includegraphics[width=0.8\linewidth]{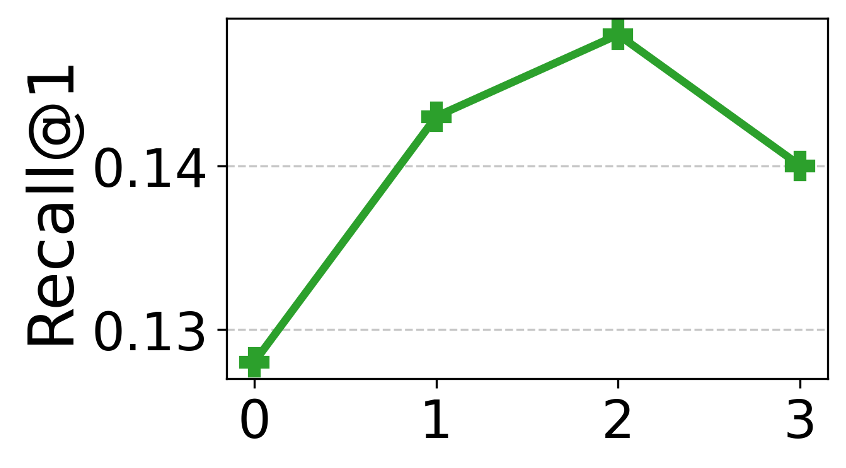}
    \Description{}
  \end{minipage}
% \vspace{-3mm}
\caption{Effect of $K_{\text{aug}}$ (left) and $K_{\text{ref}}$ (right) in $\text{KnowSA}_{\text{CKP}}$.}
\label{fig:steam_hp}
% \vspace{0.1cm}
\end{figure}

We provide analysis to guide the hyperparameter selection: the number of augmented items ($K_{\text{aug}}$) and the number of reference items ($K_{\text{ref}}$). 
Both hyperparameters should be chosen with consideration for resource constraints, such as GPU memory and the LLM's context budget.
In our experiments, the number of augmented items used is approximately 10 on average across datasets.
Figure~\ref{fig:steam_hp} reports the results for Qwen-7B on Steam, with similar trends observed across other settings.
We observe that performance improves as $K_{\text{aug}}$ increases, but saturates beyond a certain point.
This indicates that while providing supplementary information is beneficial, excessive augmentation can exceed the LLM's effective context budget.
For $K_{\text{ref}}$, a small number is generally sufficient, with the best performance achieved~at~$2$.

\section{Related Work}
\noindent
\textbf{LLM-based Recommendation.}
To improve the LLM-based recommendation~\cite{LLMRank-ECIR24, ALLMRec-KDD24, luo2024recranker}, two major strategies have emerged.
The first is model-level adaptation, which fine-tunes LLMs using interaction data.
\cite{TallRec, CoLLM23, BinLLM-ACL24, DealRec2024, BigRec2025, DLCRec2025} incorporate recommendation-specific objectives into the fine-tuning process. 
\cite{ALLMRec-KDD24} combines lightweight adapters with frozen LLMs, significantly reducing training costs.
Some methods~\cite{EXP3RT24, liu2024llm} use auxiliary data such as reviews to enhance personalization.
However, fine-tuning LLMs remains resource-intensive compared to conventional recommenders. 
Also, interaction data for fine-tuning is sparse and skewed toward popular items, limiting the model's generalization.

Another emerging direction is prompt-level adaptation, which enriches input prompts with additional information~\cite{TransRec-KDD24, EXP3RT24, LLM-TRSR-WWW24, KRagRec25}.
For example, \cite{EXP3RT24} utilizes review summaries to provide rich information for users and items, while \cite{KRagRec25} uses knowledge graphs to supplement missing knowledge.
These approaches allow LLMs to supplement incomplete knowledge without fine-tuning.
However, most existing methods adopt uniform augmentation—applying information to all items indiscriminately, which overlooks the inherent knowledge gaps in LLMs.
We propose an effective strategy to improve training-free LLM recommenders by selectively enriching items lacking sufficient knowledge.

\begin{table}[t]
\centering
% \captionsetup{skip=1pt}
\caption{Computational cost of CKP scoring. We compare the processing time (seconds) between full vs. sampled windows. Mem denotes GPU memory usage.}
\label{tab:scoring_cost}
\resizebox{\linewidth}{!}{%
\begin{tabular}{lrrcr}
\toprule
\textbf{Dataset} & \textbf{Time\textsubscript{Full}} (s) & \textbf{Time\textsubscript{Sampled}} (s) & \textbf{Speedup} & \textbf{Mem} (GB) \\
\midrule
\textbf{A-Beauty} & 903 & 395 & $2.3\times$ & 17.7 \\
\textbf{A-Gift} & 711 & 124 & $5.7\times$ & 17.8 \\
\textbf{ML-1M} & 61,362 & 602 & $101.9\times$ & 17.8 \\
\textbf{Steam} & 18,442 & 927 & $19.9\times$ & 19.2 \\
\bottomrule
\end{tabular}}
% \vspace{-0.1cm}
\end{table}

\begin{table}[t]
\centering
% \captionsetup{skip=3pt}
\caption{Efficiency comparison of different augmentation strategies on Qwen-7B. 
Overhead denotes the percentage increase in tokens and latency compared to 'No Augment'.
}
\label{tab:efficiency}
\renewcommand\theadfont{\bfseries}
\resizebox{\linewidth}{!}{%
\begin{tabular}{lcccc}
\toprule
& \multicolumn{2}{c}{\textbf{Uniform-Meta}} & \multicolumn{2}{c}{\textbf{\proposed}} \\
\cmidrule(lr){2-3} \cmidrule(lr){4-5}
\textbf{Dataset} & \thead{Tokens \\ Overhead (\%)} & \thead{Latency \\ Overhead (\%)} & \thead{Tokens \\ Overhead (\%)} & \thead{Latency \\ Overhead (\%)} \\
\midrule
\textbf{A-Beauty} & +33.0\% & +7.8\% & \textbf{+19.0\%} & \textbf{+5.6\%} \\
\textbf{A-Gift} & +87.4\% & +6.9\% & \textbf{+54.1\%} & \textbf{+5.7\%} \\
\textbf{ML-1M} & +85.6\% & +11.6\% & \textbf{+29.3\%} & \textbf{+5.5\%} \\
\textbf{Steam} & +414.3\% & +14.2\% & \textbf{+99.0\%} & \textbf{+4.1\%} \\
\bottomrule
\end{tabular}}
% \vspace{-0.4cm}
\end{table}

\noindent
\textbf{Knowledge Probing for LLMs.}
A core of our framework is to accurately estimate how much an LLM \textit{knows} about each item.
We categorize existing approaches relevant to this goal into three groups.
% PDD
First, \textit{Pre-training Data Detection (PDD)} 
aims to determine if a text was seen during pre-training~\cite{PDD-USENIX21, DCPDD-EMNLP24, Mink-ICLR24, DPDLLM-ACL24}.
PDD methods typically rely on generation likelihood, assuming that pre-seen texts will yield higher probabilities~\cite{Mink-ICLR24, DCPDD-EMNLP24, DPDLLM-ACL24}.
% UE
Second, \textit{Uncertainty Estimation (UE)} focuses on quantifying the confidence in model prediction~\cite{SurveyUE_ACL25, EigValLaplacian}.
UE methods range from latent information-based metrics (e.g., entropy) \cite{UE_entropy20, SAR_ACL24} to consistency-based measures \cite{EigValLaplacian, CCP_ACL24}; for instance, \cite{EigValLaplacian} computes the eigenvalues of a Laplacian graph constructed from the semantic similarity of sampled responses.
% Adaptive Retrieval
Third, \textit{Adaptive Retrieval} approaches~\cite{SeaKR, FLARE_EMNLP23, DRAGIN_ACL24, AdaptiveRAG_NACCL24} dynamically decide when to retrieve external documents.
Approaches include monitoring generation probabilities \cite{FLARE_EMNLP23, DRAGIN_ACL24}, leveraging internal model states \cite{SeaKR}, or employing external classifiers \cite{AdaptiveRAG_NACCL24} to predict retrieval needs.

While these approaches have proven their efficacy in general NLP tasks, they are suboptimal for recommendation.
The core issue is that they assess \textit{item-specific knowledge} in isolation, thereby overlooking the \textit{collaborative patterns}—the relationship between the user's history and the target item—essential for recommendation, as detailed in our analysis (\cref{prelim_analysis}).
To address this, we propose CKP, a new knowledge scoring method that evaluates the LLM's knowledge in a comparative setting conditioned on user interaction contexts.
This approach effectively mitigates surface-level biases and enables a more recommendation-aligned estimation of the model's actual knowledge.

\section{Conclusion}

We first provide in-depth analysis showing the necessity and challenges of addressing the knowledge gap in LLM-based recommendation.
Motivated by the analysis, we introduce \proposed, which mitigates the gap through selective knowledge augmentation.
\proposed comprises two main components: 
(1) Comparative Knowledge Probing, which estimates the LLM’s knowledge of each item, and (2) Selective Augmentation, which enriches knowledge for lesser-known items.
Extensive experiments show that \proposed consistently improves both recommendation accuracy and long-tail coverage, while also reducing the input context consumption of LLMs.
Future work may explore leveraging external sources such as knowledge graphs to further enhance augmentation.

\section*{Acknowledgments}
This work was supported by the NRF grant funded by the MSIT (No. RS-2024-00335873), the IITP grant funded by the MSIT (No.RS-2019-II191906, Artificial Intelligence Graduate School Program(POSTECH)), Korea Innovation Foundation (INNOPOLIS) grant funded by the Korea government (MSIT) (No. RS-2025-25449754).
This work was also supported by ICT Creative Consilience Program through the IITP grant funded by the MSIT (IITP-2026-RS-2020-II201819) and Basic Science Research Program through the NRF funded by the Ministry of Education (NRF-2021R1A6A1A03045425).

%%
%% The next two lines define the bibliography style to be used, and
%% the bibliography file.
\bibliographystyle{ACM-Reference-Format}
\balance
\bibliography{acmart}

%%
%% If your work has an appendix, this is the place to put it.
% \clearpage
% \appendix
% \nobalance
% \input{./sections/080_Appendix}

\end{document}